\documentclass[aps,prb,twocolumn,superscriptaddress,showpacs]{revtex4-1}
\usepackage{amsmath,graphicx,color}
\begin{document}
\title{Macroscopic description of the two-dimensional LaAlO$_3$/SrTiO$_3$ interface}

\author{V. A. Stephanovich}
\affiliation{Institute of Physics, Opole University, ul. Oleska 48, 45-052 Opole, Poland}

\author{V. K. Dugaev}
\affiliation{Department of Physics and Medical Engineering, Rzesz\'ow University of Technology,
al. Powsta\'nc\'ow Warszawy 6, 35-959 Rzesz\'ow, Poland}
\affiliation{Departamento de F\'isica and CeFEMA, Instituto Superior T\'ecnico, Universidade de Lisboa,
av. Rovisco Pais 1, 1049-001 Lisbon, Portugal}

\begin{abstract}

We propose a simple analytical model to explain possible appearance of the metallic conductivity in the two-dimensional (2D) LaAlO$_3$/SrTiO$_3$ interface. Our model considers the interface within a macroscopic approach which is usual to semiconductor heterojunctions and is based on drift-diffusion equations. The solution of these equations allows to obtain the positions of band edges as a function of distances from the interface. We show that for the 2D metallic conductivity to appear at the interface, the constituting substances should have the same type (either electronic or hole) of conductivity; in the opposite case the possible transition to metallic phase has a three-dimensional character. The results of our model calculations are in agreement with available experimental data. 

\end{abstract}

\date{\today}

\maketitle

\section{Introduction}

The oxide perovskites have an impressive range of the functional properties making them quite indispensable materials for technological applications.\cite{bib11,fcells,sze,shur,spintr} On the other hand, the experiments with these oxides posed a lot of fundamental problems in condensed matter physics. The point is that even though the bulk properties of perovskite (and those of the other symmetries) oxides are well known (see, e.g., Refs. \onlinecite{cox, imada} and references therein), the interfaces between the oxides can exhibit a multitude of very exotic properties. After the seminal paper of Ohtomo and Hwang\cite{ohtomo}, the interest to the properties of interfaces between two oxides has been renewed enourmously.\cite{okamoto} This is because the physical properties of such interfaces turn out to be much richer than those of the bulk constituents. 

Possibly, the most studied oxide interface system so far is the high-mobility confined metallic electron gas that appears in heterostructures combining two band insulators, namely, in LaAlO$_3$/SrTiO$_3$ (LAO/STO).\cite{ohtomo, pen09, kh12, th06, xie13} While the LAO and STO compounds are insulating non-magnetic oxides,  their interface can exhibit not only two-dimensional metallic conductivity\cite{ohtomo,kh12} but also the low-temperature superconductivity, ferromagnetism \cite{gang14, hwang12, brink07, kum15, lee11,lee13, har12, bell09, reyren07,cav08,shalom10, dikin11, sing09, ueno08}, their coexistence \cite{hwang12, lee11} and phase transition to insulating phase from metallic or even superconducting ones \cite{kum15, cav08, bell09, th06, cen08}. Also, anomalous magnetoresistance and Hall effect have been measured and explained \cite{seri09, zhou15, rey12}. These exotic properties of the interfaces (LAO/STO in particular) make them presently the subject of very intensive experimental and theoretical studies.

Despite continuous experimental and theoretical efforts, the physical origin of quasi-two-dimensional mobile electron gas in LAO/STO interface still remains unclear. Currently, two main scenarios coexist. First one is related to the so-called polar catastrophe model and the other one to extrinsic doping effects by the La$^{3+}$ cations, which are the $n$-type dopants in STO.\cite{tok93}  The former model can be formulated in terms of the usual electrostatics. Indeed, due to the alternating polarity of atomic layers in LAO along the [001] direction, the electrostatic potential diverges in LAO (hence the words "polar catastrophe") unless the electric charges are reconstructed at the interface. The two choices for the connection between LAO and STO impose opposite electrostatic boundary conditions. Namely, LaAlO$_3$ is composed of charged layers of (LaO)$^+$ and (AlO$_2$)$^-$, whereas the corresponding layers in SrTiO$_3$ are chargeless. Therefore, terminating the LAO on an atomic plane at the interface breaks the charge neutrality yielding the above-mentioned "polar catastrophe" at the interface. To avoid diverging interface energy, a compensating charge is required. As the LAO/STO interfacial "polar catastrophe" is quite common at the semiconductor heterointerfaces (see, e.g., Refs. \onlinecite{sze,shur}), we will be using below the spatially inhomogeneous equations, which is appropriate for the interfaces.  

We should note that in addition to above two scenarios, there is one more consisting in the formation of bulk-like oxygen vacancies in the STO layers near the interface, which provides the free carriers \cite{kal07,sie07,herranz07}. Now this scenario is considered as less probable. Nevertheless, there are still very active debates about possible origin of the observed LAO/STO interface conductivity. 

In this paper we consider a macroscopic model which makes possible to describe the 2D metallic conductivity of the LaAlO$_3$/SrTiO$_3$ interface. We consider the semi-classical \cite{sze,shur} formalism used to describe the contacts between two solids like metal-semiconductor, two semiconductors etc. This formalism, being phenomenological by its nature, utilizes the set of equations for current (resistive + diffusive) along with the Poisson and continuity equations. It allows to calculate the band bending, and, in the case when the edge of conduction band goes below the Fermi level, gives the local (in the interface) metal-insulator transition. It turns out that in the case of LAO/STO interface this formalism allows to derive simple 1D differential equations, which permit to solve this problem exactly, presenting a simple criterion of such 2D interface conductivity appearance. Moreover, we can predict that for the interface to have properly 2D conductivity, the charge carriers in constituting materials (i.e., LAO and STO in our case) should be of the same type, both electrons ($n$-$n$ junction) or both holes ($p$-$p$ junction). In the opposite case of $p$-$n$ or $n$-$p$ junctions, the resulting metallic state will be three-dimensional.

Before proceeding further, we should make a remark that in the LaAlO$_3$/SrTiO$_3$ heterostructure neither LAO nor STO are ferroelectric. The STO can become ferroelectric under an external stress and/or strain. Such stress (strain) arises due to the mismatch effects which appear inevitably at any interface, and it can be the case for LAO/STO also. On their turn, the mechanical stresses may trigger the initial band bending (see Ref. \onlinecite{yos08}, where the notched structures in the interface region have been observed experimentally) leading to electron accumulation and metallic 2D conductivity. Also, the strain effects may engender electron-phonon interaction at the interface, which in turn may generate the interfacial superconductivity. 

\section{Theoretical formalism}

\subsection{Qualitative discussion}

First of all, we consider the problem qualitatively. If we have a contact of two conducting solids in thermodynamic equilibrium, the position of Fermi level $E_F$ is the same for both of them. So far, we do not consider the formation of the Bardeen barrier at the interface (see, e.g., Refs.~\onlinecite{bardeen47, tung}), which can pin the Fermi level. We can consider the Fermi level to be "frozen", while the edges of conduction $E_c$ and/or valence $E_v$ bands are "movable". If we move the band edge $E_c$ below the Fermi level for the $n$-type of conductivity (or $E_v$ above $E_F$ for the $p$-type), the part of a sample, where the condition $E_c<E_F$ is satisfied, becomes metallic. The variation of $E_c$ and $E_v$ positions occurs due to their dependence (at wavevectors $k=0$) on spatial coordinates $x,y,z$. This dependence is formalized by the corresponding coordinate dependence of electrostatic potential $\varphi(x,y,z)$ so that the energy band bending is $e\varphi(x,y,z)$ ($e$ is the electronic charge). If there is no effects of ferroelectric domain structure (we recollect here that STO under normal conditions is nonferroelectric), there is no variation of $E_c$ and $E_v$  along $y$ and $z$ directions. So, here we consider only the important dependence on $x$ coordinate, which is perpendicular to the interface.\cite{sze,anselm} 

\begin{figure}[tbh]
\begin{center}
\includegraphics[width=0.65\columnwidth]{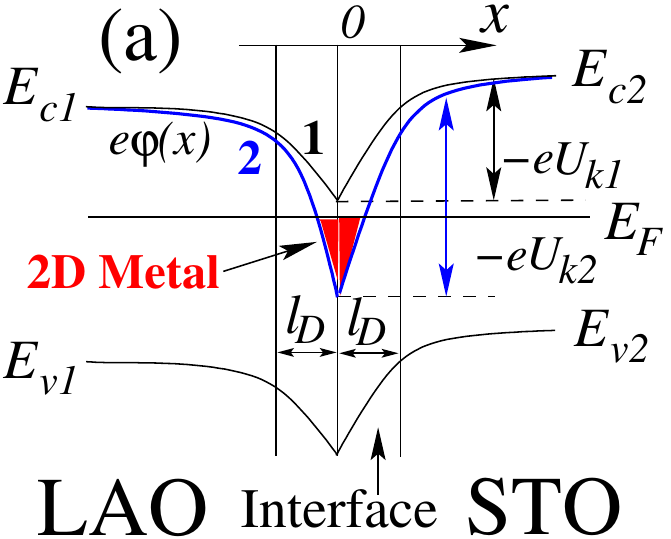} 
\end{center}
\bigskip
\bigskip
\includegraphics[width=0.65\columnwidth]{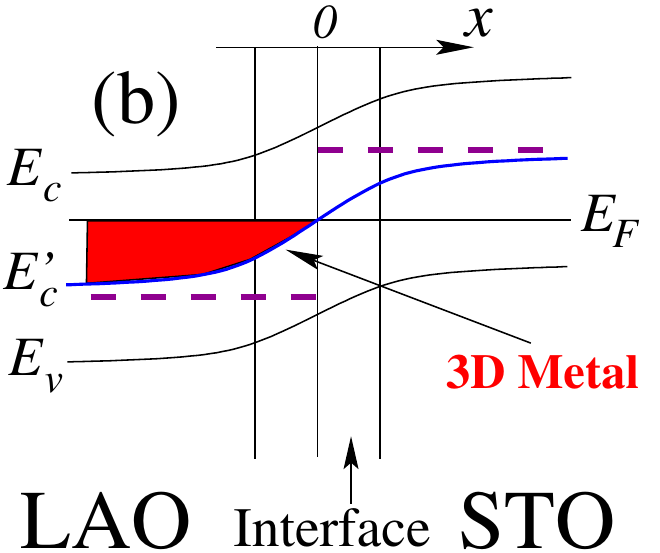}

\caption{(Color online) Panel (a). The band diagram of the LAO/STO interface, corresponding to $n$-$n$ (or $p$-$p$) junction and realizing 2D metal at the interface. The band bending is described by the electrostatic potential function $\varphi(x)$. The interface width is 2$l_D$, where $l_D$ is Debye length. $E_{c1,2}$ and $E_{v1,2}$ are the positions of the edges of the conduction and valence bands of LAO and STO respectively. As an example we consider the case of electronic conductivity, when Fermi level $E_F$ is closer to the conduction band edge. Situation 1 (black band profile curve) corresponds to dielectric interface when Schottky barrier height $e|U_{k1}|$ is less then $E_F$, while situation 2 (blue band profile curve) corresponds to metallic case $e|U_{k2}|>E_F$. The range of the interface with occupied electronic states with energies less then $E_F$ is shown by red color.  Panel (b) shows the hypothetic case of $p$-$n$ or $n$-$p$ junctions, realizing the 3D metal (full blue curve with conduction band edge $E'_c$) both at the interface and in the bulk of the one of constituent dielectrics (semiconductors). The full blue curve corresponds to the absence of interfacial charge density $\rho_s=0$, while the dashed magenta one - to $\rho_s \neq 0$, see Eq.\eqref{u4a}. The interface width and Schottky barrier heights are the same as those in panel (a). The direction of $x$ axis is shown on both panels.} 
\label{fig1}
\end{figure}

Now, if the LAO/STO interface is the contact of semiconductors of the same type of conductivity (like $n$-$n$ or $p$-$p$), the energy band bending looks like that in panel (a) of Fig.~\ref{fig1} (shown for the $n$-$n$ junction). In this case, under different Schottky barrier heights $U_k<0$, the electrostatic potential function, describing $E_c(x)$ (for the case of $n$-type conductivity) at the interface can be lower than $E_F$, thus giving the metallic conductivity near the interface. It is seen (red coloured "nib" in Fig.~\ref{fig1}a), that the region of the sample, where the metallic conductivity is realized, is confined between two blue curves $E_{c1}(x)$ and $E_{c2}(x)$ and thus is truly of 2D character as it is extended in the $yz$ plane of the sample. 

The case of $p$-$n$ (or $n$-$p$) junction is shown in Fig.~\ref{fig1}b. Now the band profile at the interface is smooth (without any "nib"), and the situation with $E'_c<$ $E_F$ is shown by the blue curve. It is seen that, contrary to the case of $p$-$p$ or $n$-$n$ junction in Fig. \ref{fig1}a, the metallic region is extended in $x$ - direction, resulting in the 3D character of metallic conductivity. Note, that while the bands on panel (a) are shown for nonzero surface charge density $\rho_s$, resulting from the presence of charged (LaO)$^+$ or (AlO$_2$)$^-$ layers at the interface, the full lines on panel (b) report the hypothetic case $\rho_s = 0$. The case of  $\rho_s \neq 0$ (see Eq. \eqref{u4a} below) is shown by the dashed lines. It is seen that in this case the conditions for 3D metal realization is even better then those for $\rho_s = 0$. This is because now 3D metal is realized in the whole rectangle, bounded by $E_F$ and $E'_c$.

This means that only the contact of semiconductors (insulators) of the same type of conductivity can generate the 2D conducting interface between them. Our quantitative analysis below will reveal the conditions (the relations between dielectric permittivities and bulk free electron concentrations in both dielectrics), under which the 2D metallic layer can emerge.     

\subsection{General expressions}

The quantitative description can be accomplished by the usual set of equations \cite{sze,shur,anselm} which are used to calculate the physical properties (like, e.g., current-voltage characteristics) of the contact between two solids. It consists of the equation for current density
\begin{equation}\label{fil1}
{\bf j}=-en\mu {\nabla} \varphi+\mu k_BT \nabla n,
\end{equation}
which includes the drift (first term) and diffusive (second term) components,
the Maxwell equation (leading to the Poisson equation for the scalar potential $\varphi $, see below)
\begin{equation}\label{fil2}
{\mathrm {div}} {\bf D}=4\pi \rho ,
\end{equation}
and the continuity equation (for generality, we present it in a non-stationary case) 
\begin{equation}\label{fil3}
\frac{\partial \rho}{\partial t}=-{\rm {div}}\, {\bf j}.
\end{equation}
The band bending spatial profile is determined by the electrostatic potential $\varphi(x)$.
As we want to calculate the band profile, we consider the case of ${\bf j}=0$, when, due to the time independence, Eq.~\eqref{fil3} is satisfied identically. In Eqs.~\eqref{fil1} to \eqref{fil3}, $\mu$ and $n$ are, respectively, the mobility and concentration of carriers, $T$ is the temperature, $k_B$ is the Boltzmann constant, and $\varphi (x)$ is the electrostatic potential. The first term in Eq.~\eqref{fil1} is the resistive (drift) current ${\bf j}_d=\sigma {\bf E}$, where $\sigma=en\mu$ is the conductivity and ${\bf E}=-\nabla \varphi$ is electric field, while the second term is the diffusion current, proportional to the carriers concentration gradient. By $\rho$ we denote the charge density, 
$\rho=e(N_d-N_a-n)$, where 
$N_d$ and $N_a$ are, respectively, the concentrations of donors and acceptors. 

In the general case, $N_d$, $N_a$ and $n$ can be some functions of coordinates. Here we consider case of an intrinsic semiconductor with $N_d=$ $N_a=0$ and
\begin{equation}\label{fil5}
\rho=e(n_0-n),
\end{equation}
where $n_0$ is a constant charge carrier density (for definiteness we speak about electrons) in the bulk of a substance, where $\rho=0$. 

With respect to the ordinary relation ${\bf D}=$ $\varepsilon {\bf E}$ (where ${\bf D}$ and ${\bf E}$ are the electrical displacement and electric field vectors, respectively, and $\varepsilon$ is the dielectric permittivity of a dielectric), the Maxwell equation \eqref{fil2} transforms into Poisson's equation

\begin{equation}\label{fil6}
\Delta \varphi = \frac{4\pi e}{\varepsilon}(n_0-n).
\end{equation}

In the 1D version of this model, all the gradients become the derivatives with respect to $x$ only.
\begin{eqnarray}
&& j\equiv j_x=-en(x)\mu\frac{d\varphi}{dx}+\mu k_BT\frac{dn}{dx},  \label{fil7a} \\
&&\frac{d^2\varphi}{dx^2}=\frac{4\pi e}{\varepsilon}(n-n_0). \label{fil7} 
\end{eqnarray}

To obtain the band bending function $\varphi(x)$ for our model, we first put $j=0$ in \eqref{fil7a}, thus obtaining the relation between functions $n(x)$ and $\varphi(x)$. We have   
\begin{equation}\label{tin1}
n(x)=n_ke^{\frac{e\varphi(x)}{k_BT}},
\end{equation}
where we choose the arbitrary constant $n_k$ to be the electron concentration at the interface: $n_k=n(x=0)$, \cite{anselm}
\begin{equation}\label{tin2}
n_k=n_0e^{-\frac{eU_k}{k_BT}},
\end{equation}
where $U_k$ is Schottky barrier height.
Combining the expressions \eqref{tin1} and \eqref{tin2}, we obtain
\begin{equation}\label{tin3}
n(x)=n_0\exp\left[\frac{e(\varphi(x)-U_k)}{k_BT}\right],
\end{equation}
which yields finally the equation for the band bending function $\varphi(x)$

\begin{equation}\label{tin4}
\frac{d^2\varphi}{dx^2}=\frac{4\pi e}{\varepsilon}n_0\left\{\exp\left[\frac{e(\varphi(x)-U_k)}{k_BT}\right]-1 \right\}. 
\end{equation}

To account for the interface between dielectrics 1 and 2, we should consider two equations for the potentials $\varphi_{1,2}(x)$ and augment them by the boundary conditions\cite{land8} at $x=0$
\begin{eqnarray}
&&D_{1n}-D_{2n}=-4\pi \rho _s,  \label{u4a} \\
&&E_{1t}=E_{2t},  \label{u4b}
\end{eqnarray}
where $\rho _s$ is a surface charge density at the interface; the indices $n$ and $t$ stand for normal (to the interface, i.e., in the $x$ direction) and transversal directions, respectively. For definiteness, we consider here dielectric 1 to be the left one (LAO in the Fig. 1) and 2 to be the right one (STO in the Fig. 1). 

Hence, here we consider a possibility for the electric charge to accumulate at the interface between two insulators. This is the key point of our modelling \cite{note} since, as we have mentioned above, the metallic interfacial conductivity appears due to surface charge reconstruction, needed to avoid the "polar catastrophe". This charge, in turn, should be initially accumulated in the interface by means of stacking of either (LaO)$^+$ or (AlO$_2)^-$ layers over neutral SrO or TiO layers. The latter fact is reflected in our macroscopic model by introduction of surface charge density $\rho_s$.
Indeed, without interfacial electric charges, Eq.~\eqref{u4a} leads to continuity of the normal component of induction, $\varepsilon _1\varphi '_1(0)=\varepsilon _2\varphi '_2(0)$, and, accordingly, to the monotonic dependence of potential $\varphi (x)$ for all $x$. On the other hand, for $\rho _s\ne 0$ and relatively small contact potential difference, $|\varphi (\infty)-\varphi (-\infty )|/l_D\ll |\varphi '(0)|$, the above surface charge $\rho _s$ acts in a way that $D_{1n}=-D_{2n}$ at the interface. This leads to non-monotonic behaviour of $\varphi (x)$.  

In this case the condition \eqref{u4b} with respect to relation $E(x)=-\varphi '(x)$ gives 
\begin{equation}\label{tin5}
\varphi_1(0)=\varphi_2(0)\equiv \varphi_0,
\end{equation}
while \eqref{u4a} yields
\begin{equation}\label{tin6}
\varepsilon_1\, |\varphi'_1(0)|=\varepsilon_2\, |\varphi'_2(0)|
\end{equation}
with different signs of derivatives $\varphi '_1(0)$ and $\varphi '_2(0)$
(as shown schematically in Fig.~\ref{fig1},a).
 
One more boundary condition, imposed by the physics of the problem, stems from the fact that $n_{1,2}(|x| \to \infty)=n_{01,2}$. For the potential this yields (see Eq. \eqref{tin3})
\begin{equation}\label{tin6a}
\varphi_{1,2}(|x| \to \infty)=U_{k1,2}.
\end{equation}

The pair of Poisson equations \eqref{tin4} for potentials $\varphi_1$ and $\varphi_2$ along with boundary conditions \eqref{tin5}, \eqref{tin6} and \eqref{tin6a} constitute the formalism which permits to investigate our problem quantitatively.  

\subsection{Linear theory}

The usual approach in the theories of metal-dielectric contact,\cite{sze, shur, anselm} is to assume the argument of exponent in Eq. \eqref{tin4} to be small so that we obtain from \eqref{tin4} two equations for $\varphi_1$ and $\varphi_2$
\begin{equation}\label{tin7}
\frac{d^2\varphi_{1,2}}{dx^2}=\frac{\varphi_{1,2}(x)-U_{k1,2}}{l^2_{D1,2}},\ l^2_{D1,2}=
\frac{\varepsilon_{1,2}k_BT}{4\pi e^2n_{01,2}}, 
\end{equation}
where $l_{D1,2}$ is a Debye screening lengths in the dielectric 1 or 2. It had been shown \cite{sze, anselm} that the linear approach gives qualitatively the same results as the initial nonlinear one \eqref{tin4} but is much easier to solve. 
 
If we consider the $n$-$n$ or $p$-$p$ junction in the framework of linear approximation \eqref{tin7}, we have function $\varphi(x)$ to be even function, see Fig. 1a. Our analysis shows that in this case we can consider only positive half-axis $0< x <\infty$, obtaining the solution for negative half-axis by simply changing the sign in the corresponding exponent. Then we have
\begin{equation}\label{tin8}
\varphi_1=U_{k1}+C_1e^{-x/l_{D1}},\ \varphi_2=U_{k2}+C_2e^{-x/l_{D2}}.
\end{equation}
With respect to conditions \eqref{tin5} and \eqref{tin6} and introducing the dimensionless parameter
\begin{equation}\label{tin10}
S=\frac{\varepsilon_2l_{D1}}{\varepsilon_1l_{D2}}=\sqrt{\frac{\varepsilon_2n_{02}}{\varepsilon_1n_{01}}}, 
\end{equation}
we obtain the following explicit form of solutions of equaltion \eqref{tin7}
\begin{eqnarray}
\varphi_1&=&U_{k1}+\frac{S(U_{k2}-U_{k1})}{S-1}e^{-x/l_{D1}},\nonumber \\ 
\varphi_2&=&U_{k2}+\frac{U_{k2}-U_{k1}}{S-1}e^{-x/l_{D2}}. \label{tin11}
\end{eqnarray}

Now we are in the place to derive the criterion of existence of the metallic interface. The corresponding criterion can be easily seen from Fig.~\ref{fig1}a and yields
\begin{equation}\label{tin13}
\frac{E_F}{e}>\varphi_0=\frac{U_{k1}-SU_{k2}}{1-S}.
\end{equation}

The criterion \eqref{tin13} explicitly relates the parameters of the materials 1 and 2 (their dielectric permittivities, bulk electronic concentration and Schottky barrier heights) to the Fermi level of the resulting structure.

We now make some numerical estimations based on the condition \eqref{tin13}. The parameters of the interface can be taken from Ref. \onlinecite{mor2}. Dielectric permittivities $\varepsilon_{LAO}=$ $\varepsilon_1=$ 28, $\varepsilon_{STO}=$ $\varepsilon_2=$ 43. The bulk electronic concentrations  $n_{01} \sim $ $n_{02}=$ $10^{11} - 10^{12}$ cm$^{-3}$.  The Schottky barrier heights have been estimated to be $U_{k1,2}\sim $0.3-0.5~eV. Taking $n_{02}/n_{01}=10$, we obtain $S=3.919$ and for $U_{k1}=0.52$ eV,  $U_{k2}=0.26$ eV $\varphi_0=0.171$~eV, which is certainly less then $E_F \sim $1~eV. We can see that for the typical parameters of the LAO/STO structure the criterion \eqref{tin13} can be fulfilled. 

Having the results of linear model, it is instructive now to consider the nonlinear case \eqref{tin4} and see how the criterion \eqref{tin13} changes in it.  

\subsection{Nonlinear theory}

As we have mentioned above, the qualitative results of nonlinear model \eqref{tin4} are exactly the same as those of linear one. It is instructive, however, to present the complete solution of nonlinear model  \eqref{tin4} and compare it graphically with the solution of linear one. 
The nonlinear differential equation \eqref{tin4} can be solved in quadratures by means of substitution 
$d\varphi/dx=f(\varphi)$. In this case the first integral of Eq. \eqref{tin4} assumes the following form (for a moment we suppress the indices 1,2)
\begin{equation}\label{nl2}
\frac 12\left(\frac{d\varphi}{dx}\right)^2+C=\frac{4\pi e n_0}{\varepsilon}\left\{\frac{k_BT}{e}\exp\left[\frac{e(\varphi-U_k)}{k_BT}\right]-\varphi \right\}. 
\end{equation}
The constant $C$ can be found from the boundary condition \eqref{tin6a}, which implies that  
\begin{equation}\label{nl3}
\left.\frac{d\varphi_{1,2}}{dx}\right|_{x \to \pm \infty}=0.
\end{equation}

In this case the solution of Eq. \eqref{tin4} in quadratures assumes the form
\begin{eqnarray}
\int_{\varphi_0}^\varphi \frac{dz}{\sqrt{e^\psi-1-\psi}}=\pm x\sqrt{2}\sqrt{\frac{4\pi n_0k_BT}{\varepsilon}},\label{nl6} \\
\psi=\frac{e(z-U_k)}{k_BT},\ \varphi \leq U_k. \nonumber 
\end{eqnarray}  
In the expression \eqref{nl6} we account for the fact that $\varphi (x=0)=\varphi_0$. The determination of yet unknown value $\varphi_0$ can be accomplished utilizing boundary condition \eqref{tin6}. 
\begin{eqnarray}
f\left[\frac{e(\varphi_0-U_{k1})}{k_BT}\right]=Sf\left[\frac{e(\varphi_0-U_{k2})}{k_BT}\right], \nonumber \\
f(x)=e^x-1-x, \label{nl7}
\end{eqnarray}   
where $S$ is defined by the expression \eqref{tin10}.

\begin{figure}[tbh]
\centering
\includegraphics[width=\columnwidth]{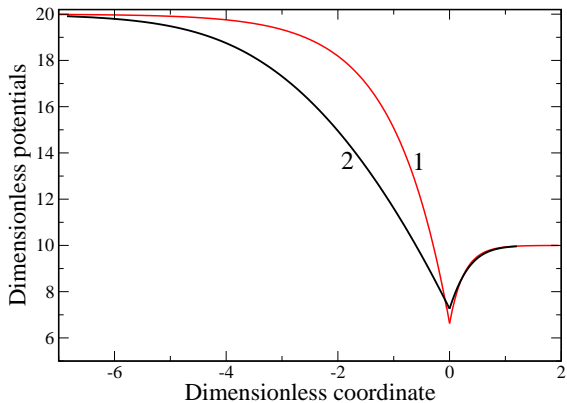}
\caption{(Color online) The plots of dimensionless potentials $z_{1,2}=e\varphi_{1,2}/k_BT$ versus dimensionless coordinate $x_1=x/l_{D1}$. Curve 1 corresponds to linear case \eqref{tin11}, curve 2 - to nonlinear \eqref{nl10}. For the linear case, the dimensionless value $e\varphi_0/k_BT$ is $z_{0\, lin}=6.615$. For both cases and above materials parameters $S_1=3.919$.} 
\label{fig2}
\end{figure}

The equation \eqref{nl7} is a transcendental equation for determination of $\varphi_0$ for known values 
of $U_{k1,2}$, $\varepsilon_{01,2}$ and $n_{01,2}$. We note that expansion of exponents in \eqref{nl7} in power series at small arguments generates the equation \eqref{tin13} from linear theory. Also, if we expand the exponential under square root in \eqref{nl6}, we obtain the solution \eqref{tin11} of linear problem. 

Now, in nonlinear theory, the "metallicity criterion" has the same form \eqref{tin13} but the equation for determination of $\varphi_0$ \eqref{nl7} is transcendental. 

To plot the dependence $\varphi(x)$ on the base of Eq. \eqref{nl6} we use following dimensionless variables: 
\begin{eqnarray}
&&\frac{e(z-U_{k1,2})}{k_BT}=\eta,\ z_{0,1,2}=\frac{e\varphi_{0,1,2}}{k_BT}, \
y_{1,2}=\frac{eU_{k1,2}}{k_BT},\nonumber \\
&&x_1=\frac {x}{l_{D1}},\ S_1=\frac{l_{D1}}{l_{D2}}=\sqrt{\frac{\varepsilon_1n_{02}}{\varepsilon_2n_{01}}}. \label{nl8}
\end{eqnarray}

In the dimensionless variables \eqref{nl8}, the pair of equations \eqref{nl6} assumes the form (we now restore the indices 1,2)
\begin{eqnarray} 
&&\int_{z_0-y_1}^{z_1-y_1}\frac{d\eta}{\sqrt{e^\eta-1-\eta}}=-\sqrt{2}x_1,\nonumber \\ \nonumber \\
&&\int_{z_0-y_2}^{z_2-y_2}\frac{d\eta}{\sqrt{e^\eta-1-\eta}}=-\sqrt{2}S_1x_1. \label{nl10}
\end{eqnarray}

The numerical solution of the set \eqref{nl10} occurs as follows. First, we determine $z_0$ (dimensionless $\varphi_0$) from the equation \eqref{nl7}. Then we substitute them into the integrals \eqref{nl10}, which are to be calculated numerically for each value of $z_1<y_1$ and $z_2<y_2$. The resulting inverse functions $z_{1,2}(x_1)$ are the desired potentials coordinate dependences in dimensionless units \eqref{nl8}.

For the parameters of above linear theory $U_{k1}=0.52$~eV,  $U_{k2}=0.26$~eV and room temperature $T=300$~K we have $y_1=20$ and $y_2=10$. The solution of transcendental equation \eqref{nl7} gives in this case $z_0=7.2621$, which in dimensional units correspond to $\varphi_0=0.188$~eV, which is close to linear value $\varphi_{0\, lin}=0.171$~eV. The plots of $z_{1,2}(x_1)$ for linear and nonlinear case are reported in Fig.~2. It is seen, first of all, that the linear case captures the situation qualitatively pretty good (note that the same is true for $p$-$n$ junctions) and for presented materials parameters gives also good quantitative coincidence with nonlinear case. This means, that in the problem of LAO/STO interface we can safely use linear approximation and explicit "metallicity criterion" \eqref{tin8}. We finally note here that varying the parameters of substances 1 and 2 (like their dielectric permittivities and bulk electronic concentrations) as well as temperature, we can obtain "better metallicity" in the interface. Also, the presented formalism permits to calculate current-voltage characteristics of the interface as well as time-dependent (nonequilibrium like frequency-dependent conductivity) quantities. Latter calculations, probably, can be done only numerically.

\section{Discussion and conclusions}

As there is still no complete description of the LAO/STO interface ionic and electronic structure, our phenomenological model elucidates the conditions for the interface to become conductive.  
Fig. \ref{fig1} can be well regarded as a band diagram of the LAO/STO heterointerface. For this interface to become metallic, we should "pull" the conduction (for $n$-type of conductivity) band edge below Fermi level. This fact is formalised by the criterion \eqref{tin13}, augmented by Eq. \eqref{nl7} in nonlinear case. 

As we have discussed above, our calculations are in agreement with electrostatic, "polar catastrophe" scenario of emergence of metallic interface. Although this scenario has its shortcomings \cite{bib11,spintr} (like the discrepancy between the expected surface carrier density $n_s= 3.5 \times 10^{14}$~cm$^{-2}$ and the measured values around 10$^{13}$ cm$^{-2}$), the experimental investigations of LAO/STO interface by means of {\em {in situ}} photoemission spectroscopy \cite{yos08} show that this scenario is pretty realizable by means of formation of the notched structure (where charge carriers can accumulate) inside the 
nonpolar STO layers. This accumulation is necessary to avoid the "polar catastrophe" (the electric potential divergence) in the polar LAO layers.  In other words, the divergence can be lifted by the formation of the long-range electric potential inside the STO whose spatial variation is governed by the band bending, formed in turn by the charge carriers in the STO layers. Our theory takes this effect into account. The other question is the origin of these excessive charge carriers (the electrons for concreteness). In principle, these carriers can be supplied either by the above discussed oxygen vacancies \cite{kal07,sie07,herranz07} or extrinsic doping in STO.  However, widely accepted explanation of the source of carriers is also within the "polar catastrophe" model, namely by the charge transfer from the LAO to the STO of 1/2 electron per unit cell to avoid the electrostatic potential divergence due to the stacking of a polar material onto a non-polar one. In the spirit of that, our model makes the important physical conclusion that truly 2D metallic interface is possible if both constituents have the same ($n$-$n$ to $p$-$p$) types of conductivity. 

Note, that the important peculiarity of the LAO/STO interface is the strong sensitivity of its transport properties to external electric field that permits to tune its physical properties by that field,
which in turn, advances considerably its practical applications.  

We recollect finally the main conclusions of our theory. To realize the 2D conducting interface between two dielectrics (LAO and STO for instance), we need the fulfilment of two conditions - the same ($n$-$n$ or $p$-$p$) type of conductivity of two interface forming dielectrics and the filling of "Fermi sea" by the free electrons or holes. Latter condition \eqref{tin13} states simply that (conduction for electrons and valence for holes) band edge $e\varphi_0$ should lie below the Fermi level of a structure. In nonlinear case the $\varphi_0$ is to be found from transcendental equation \eqref{nl7}. 

\begin{acknowledgments}
This work was supported by the National Science Center in Poland as a research project
No.~DEC-2012/06/M/ST3/00042.
\end{acknowledgments}


\begin{references}
\bibitem{note}
We believe that our assumption (made within the macroscopic electrodynamics model) about accumulated interfacial charge is in agreement with some earlier microscopic considerations of the interface vacancies, possible mixed valence state of Ti atoms, or interfacial reconstruction.
\end{references}
\end{document}